\documentclass[aip,amsmath,amssymb]{revtex4-1}

\usepackage{graphicx}
\usepackage{dcolumn}
\usepackage{bm}

\usepackage[utf8]{inputenc}
\usepackage[T1]{fontenc}
\usepackage{mathptmx}
\usepackage{etoolbox}
\usepackage{braket}
\usepackage{xcolor}
\usepackage{comment}
\usepackage{tikz}

\makeatletter
\def\@email#1#2{%
 \endgroup
 \patchcmd{\titleblock@produce}
  {\frontmatter@RRAPformat}
  {\frontmatter@RRAPformat{\produce@RRAP{*#1\href{mailto:#2}{#2}}}\frontmatter@RRAPformat}
  {}{}
}%
\makeatother
\begin{document}

\title{Quantum routing of information using chiral quantum walks}

\author{Alberto Bottarelli}
 \email{alberto.bottarelli@unitn.it.}
 \affiliation{ 
Pitaevskii Center for Bose-Einstein Condensation and Department of Physics, University of Trento, Via Sommarive 14, Povo(TN)}
 
\author{MassimoFrigerio}\email{massimo.frigerio@unimi.it}
\affiliation{Quantum Technoloy Lab \& Quantum Mechanics Group,   Department of Physics "A. Pontremoli", Università degli Studi di Milano,
I-20133 Milano, Italy}%
 \affiliation{Istituto Nazionale di Fisica Nucleare, Sezione di Milano, I-20133 Milano, Italy}%

\author{Matteo G. A. Paris}%
 \email{matteo.paris@fisica.unimi.it}
\affiliation{Quantum Technoloy Lab \& Quantum Mechanics Group,   Department of Physics "A. Pontremoli", Università degli Studi di Milano,
I-20133 Milano, Italy}%
 \affiliation{Istituto Nazionale di Fisica Nucleare, Sezione di Milano, I-20133 Milano, Italy}%

\date{\today}

\begin{abstract}

We address routing of classical and quantum information over quantum network, and 
show how to exploit chirality to achieve nearly optimal and robust transport. In particular, we prove how continuous time chiral quantum walks over a minimal graph 
is able to model directional transfer of information over a network. At first, we 
show how classical information, encoded onto an excitation localized at one vertex 
of a simple graph, may be sent to any other chosen location with nearly unit fidelity 
by tuning a single phase. Then, we prove that high-fidelity transport is also possible 
for coherent superpositions of states, i.e. for routing of quantum information.
Furthermore, we show that by tuning the phase parameter one obtains universal quantum routing, i.e. indipendent on 
the input state. In our scheme, chirality is governed by a single phase, and the routing probability is robust against fluctuations of this parameter. Finally, we address characterization of quantum routers and show how to exploit 
the self energies of the graph to achieve high precision in estimating the phase parameter.
\end{abstract}

\maketitle
\section{Introduction}
With the advent of near-future quantum technologies, the problem of transport and routing of energy and quantum information on a network is quickly becoming crucial. In particular, whenever systems with quantum properties are employed to store information locally, its transfer between separate places and its routing towards different target locations is a nontrivial problem. A common strategy is to couple the systems to a field (typically photons) and to use the field as a mediator. However, these protocols may have substantial drawbacks with respect to noise sensitivity and fidelity of the transferred quantum state. Moreover, whenever a large number of transfers is required, the impossibility of amplifying the quantum signal due to the no-cloning theorem makes these issues even more detrimental. 

As a matter of fact, quantum spin chains provide a viable alternative for the transport of quantum information on a finite 1D lattice\cite{bose2003quantum}. Since this aspect is fully captured in the single excitation subspace of these systems, a fully equivalent description can be provided in the framework of continuous-time quantum 
walks on graphs\cite{mulken2011continuous,venegas2012quantum,portugal2013quantum}. 
However, conventional quantum walks do not allow for directional transport of the excitation because their Hamiltonian generators are real matrices. In order to 
overcome these limits, one may employ chiral quantum walks with complex Hamiltonians, 
which stand out as the simplest generalisation to introduce routing in a quantum walks framework
\cite{zimboras2013quantum,lu2016chiral,Sett2019,frigerio2021generalized,Khalique2021,frigerio2022quantum}.

In this paper, we address the problem of optimally routing an excitation between two different possible target sites on the simplest graph apt to this task with chiral quantum walks. We make connections with spin chains and we also provide an alternative way to transport quantum information exploiting just the degrees of freedom of the quantum walk. This allows us to significantly reduce the dimensionality of the underlying Hilbert space with respect to a spin chain, thus improving the resilience to decoherence. We also address the robustness of transport with respect to optimal values of the phase, and analyze in some detail the characterization of the router i.e. the problem of estimating the value of the phase by quantum probing.  

The paper is structured as follows: In Section II, we establish notation and recall some elements of the theory  of continuous-time chiral quantum walks. In Sec III, we introduce the system studied in this work, 
and present our results about routing of classical and quantum information. Section IV is devoted to the characterization of the quantum router, whereas Section V closes the paper with some concluding remarks. 
\section{Continuous-time Chiral Quantum Walks}
Let us start by introducing the notation that will be used during the rest of the work. Starting from a connected simple graph G(V,E) where $ V=\{1,...,N\} $ with $N=|V|$ specifies the vertices and $E$ the connections, we define the site basis $\{ \ket{1},...,\ket{N}\}$ such that $\braket{i|j}=\delta_{ij}$ , $i,j=1,...N$. Each element $\ket{i}$ of the basis represents a state localized on the vertex  $i$. Given an initial state $\ket{\psi}$, the CTQW is defined as the time evolution of the state through a unitary operator $U=e^{-iHt}$, where $H$ is a suitable Hamiltonian that satisfies the topology of the graph, usually assumed to be the Laplacian or the adjacency matrix of the graph. Those choices, however, do not allow for directional transport of the excitation because the Hamiltonian is a real matrix and any transition amplitude $\langle|e^{- i H t}|k\rangle$ is symmetric under the exchange $j \leftrightarrow k$. In order to 
overcome these limits, and discuss directionality of quantum transport, one should assume a Hermitian complex Hamiltonian corresponding to chiral quantum walks, which themselves stand out as the simplest generalisation to introduce routing in a quantum walks framework \cite{frigerio2021generalized}.

Earlier works about chiral quantum walks \cite{lu2016chiral,zimboras2013quantum} assumed some specific 
form of chiral Hamiltonians, without referring to physical systems where such interactions may take place. 
A systematic approach to chiral quantum walks has been put forward in \cite{frigerio2021generalized}, introducing a full characterization of all the possible Hamiltonians describing the time evolution over a given topology. Those Hamiltonians are the ones satisfying $L_{ij}=H^2_{ii}\delta_{ij}-|H_{ij}|^2$, where $L$ is the Laplacian of the graph. This equation has an infinite number of solutions (i.e. of valid Hamiltonians), which are  of the form 
    \begin{equation}
        H=\sum_{j\neq k\in E} \left[ e^{i\theta_{jk}}\ket{j}\bra{k}+ e^{-i\theta_{jk}}\ket{k}\bra{j}\right] + \sum_{s=1}^{N} \gamma_s \ket{s}\bra{s}\label{genH}
    \end{equation}
where the off-diagonal phases $\theta_{jk}$ and the diagonal self-energies $\gamma_s$ are free parameters not determined by the sole topology.

\subsection{A remark on the number of independent phases}
Among the infinite number of Hamiltonians that are compatible with a given topology, those with a given set of self-energies $\left\{\gamma_s\right\}$ may be transformed one to each other by unitary trasformations 
$U$ that acts diagonally on the site basis, i.e. 
\begin{equation}\label{urep}
U\ket{j}=e^{i\phi_j}\ket{j} \quad j=1,...,N\,.
\end{equation}
Applying a transformation of this kind to a given Hamiltonian, the off-diagonal phases $\theta_{jk}$ are mapped to $\theta'_{jk}=\theta_{jk}+(\phi_j-\phi_k)$. This class of transformations changes the transition amplitudes from a site to another, but do not affect the site-to-site transition probabilities $|\braket{j|e^{-iHt}|k}|^2$, as it may be easily seen from
 \begin{align}
    |\bra{j}e^{-iH't}\ket{k}|^2
    =|\bra{j}U^\dag e^{-iHt} U\ket{k}|^2 = 
    |e^{i(\phi_j-\phi_k)}\bra{j}e^{-iHt}\ket{k}|^2=
    \bra{j}e^{-iHt}\ket{k}|^2\,.
    \end{align}    
In other words, if one is interest only in the site-to-site transition probabilities, the set of possible Hamiltonians naturally splits into equivalence classes, and one may employ the most convenient representative by re-phasing sites via a unitary transformation of the form (\ref{urep}). This also means that if the quantities of interest are the site-to-site transition probabilities, the number of independent parameters (phases) at disposal to engineer the dynamics is smaller than the number of off-diagonal elements.

Let us now introduce the total phase along a loop of the graph as $\Theta \equiv \sum_{j\neq k} \theta_{jk}$, where the sum is performed over the links defining the loop. Using Eq. (\ref{urep}) it is easy to see 
that this quantity is invariant under site re-phasing, i.e. $\Theta'=\sum_{j\neq k}\theta'_{jk}=\Theta$. This means that for graphs with no loops, e.g. any tree graph, the transition probabilities cannot be changed using complex Hamiltonians (since all the off-diagonal phases may be removed by re-phasing). On the other hand, in graphs with loops, the number of free phases at disposal to engineer the transition probabilities (e.g. to obtain directionality) is equal to the number of loops. As we will see in the following Section, chiral effects induced by off-diagonal phases are enough to obtain quantum routing of information even in graphs with a single loop.

\section{The Quantum Router}
Routing is a fundamental function of any network, classical or quantum. Indeed, networks where information travels across different devices has to be equipped with a means of selecting paths through the network itself \cite{Pant2019,9946408}.

\begin{figure}[h!]
\centering
\includegraphics[width=0.5\textwidth]{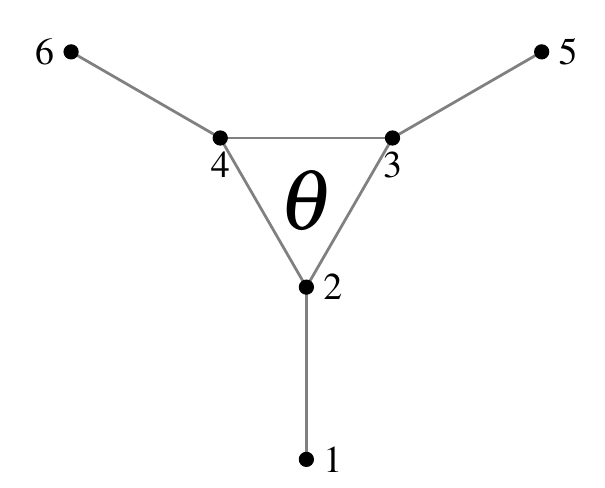}
\caption{\label{fig::router} Schematic diagram of a chirality-based minimal-topology quantum router. Vertex labels are those used in the main text. The overall Hamiltonian of the quantum walker is parametrized by the total phase along the loop, denoted by $\theta$, and by the difference $\gamma \equiv \gamma_{int}-\gamma_{ext}$ among the self-energies of the internal and external links.}
\end{figure}

The goal of thi s section is to show how a nearly optimal {\em quantum router} may be built using 
chiral quantum walk on a simple graph. In particular, we consider the minimal structure depicted in Fig \ref{fig::router} and show that by tuning the single parameter $\theta$ one may route an excitation 
initially located at site $|1\rangle$ to site $|5\rangle$ or $|6\rangle$ with high probability. 
This task is not possible with standard (i.e. non chiral) quantum walk, whereas exploiting chirality one may achieve transport with probability very close to unit. In addition, we show that routing and nearly optimal transport are also possibile for an excitation initially prepared in superposition, i.e. any superposition of the form
$|\psi\rangle\propto\ket{1}+e^{i\phi}\ket{2}$ may be sent with high fidelity to 
$|\psi\rangle\propto\ket{5}+e^{i\phi}\ket{3}$ 
or to  $|\psi\rangle\propto\ \ket{6}+e^{i\phi}\ket{4}$
by tuning the phase $\theta$.

The schematic diagram of our chirality-based minimal-topology quantum router is shown in Fig \ref{fig::router}. Vertex labels are those used throughout the paper. Among the possible {\em bona fide} chiral Hamiltonians of the form  (\ref{genH}) we focus on those with 
$\gamma_j=\gamma_{int}$ for $j=2,3,4$, 
$\gamma_j=\gamma_{ext}$ for $j=1,5,6$,
$\theta_{ij}=0$ for the links outside the loop, and $\theta_{ij}=\frac{\theta}{3}$ for links 
along the loop. With this choice we ensure that the rotational symmetry of the graph is preserved and  that the only relevant phase is the total phase $\Theta\equiv\theta$ along the loop. These features make the structure in Fig \ref{fig::router} an ideal candidate to route quantum information and, at the same time, maintain the problem analytically treatable.

We now notice that the site-to-site transition probabilities do not change if we consider a modified Hamiltonian where a term proportional to the identity, e.g. $\gamma_{ext}\mathrm{I}$, is added (subtracted). This reveals that 
only the difference $\gamma \equiv \gamma_{int}-\gamma_{ext}$ is relevant for engineering the dynamics. We thus consider the effective Hamiltonian $H \rightarrow H - \gamma_{ext}\mathrm{I}$.
The explicit form is given by $H=\ket{1}\bra{2}+ e^{i\frac{\theta}{3}}\ket{2}\bra{3}+e^{-i\frac{\theta}{3}}\ket{2}\bra{4}+e^{i\frac{\theta}{3}}\ket{3}\bra{4}+\ket{3}\bra{5}+\ket{4}\bra{6} + h.c. + \gamma \left(\ket{2}\bra{2}+\ket{3}\bra{3}+\ket{3}\bra{4}\right)$ or, in matrix form, by

\begin{equation} \label{eq::RouterHamiltonian}
    H=
    \left[{\begin{array}{cccccc}
         0 & 1 & 0 & 0 & 0 & 0  \\
         1 & \gamma & e^{-i\frac{\theta}{3}} & e^{i\frac{\theta}{3}} & 0 & 0  \\
         0 & e^{i\frac{\theta}{3}} & \gamma & e^{-i\frac{\theta}{3}} & 1 & 0  \\
         0 & e^{-i\frac{\theta}{3}} & e^{i\frac{\theta}{3}} & \gamma & 0 & 1  \\
         0 & 0 & 1 & 0 & 0 & 0  \\
         0 & 0 & 0 & 1 & 0 & 0  \\
    \end{array}}\right]
\end{equation}

\begin{figure}[h!] 
	\centering
	\includegraphics[width=0.95\textwidth]{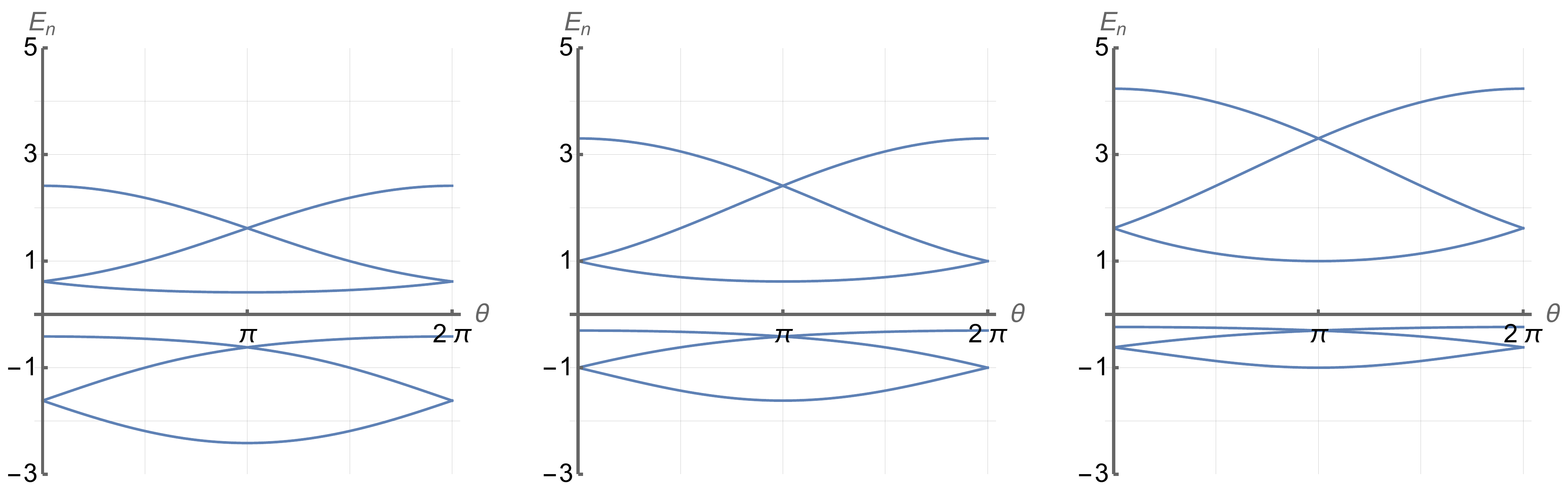}
	\caption{\label{f:eiv} Eigenvalues of H as function of $\theta$ for $\gamma = 0, 1,2$ (from left to right).}
\end{figure}

This Hamiltonian reflects the rotational symmetry of the graph, and we may thus assume that its eigenstates 
satisfy the relations 
\begin{align}\label{rsym}
\braket{E_n|k_{int}} =e^{\frac{2\pi k (n-1)}{3}}\,\alpha \qquad
\braket{E_n|k_{ext}}= e^{\frac{2\pi k (n-1)}{3}}\,\beta
\end{align}
where $\alpha$ and $\beta$ are real numbers, $n=1,2,3$, and $\ket{k}$ with $k_{int}=2,3,4$, $k_{ext}=1,6,5$ 
denote localized states. Using this ansatz, the eigenvalues of the Hamiltonian are easily found as 
\begin{align}\label{eq::eigenvalues}
          E_n & = S_n+(-1)^{\hbox{\footnotesize mod}(n,2)}\sqrt{1+S^2_n} \\
          S_n & = \frac{\gamma}{2} + \cos \left[\frac{\theta}{3} + \frac{4 \pi}{3}\hbox{mod}(n-1,3)\right]\,.
    \end{align}
The eigenvalues are shown as a function of $\theta$ in Fig. \ref{f:eiv} for $\gamma=0, 1, 2$. For increasing $
\gamma$, the three negative eigenvalues tend to coalesce and vanish as $-1/\gamma$, whereas the three positive ones increase their magnitudes. For negative values of $\gamma$ the behaviour of negative and positive eigenvalues is reversed. 

The corresponding eigenvectors are given by
    \begin{equation}
    \label{eq:eigenv}
        \ket{E_n}=\frac{1}{\sqrt{3(1+E_n^2)}}\\
            \begin{pmatrix}
            1\\
            E_n\\
            e^{i\frac{2\pi (n-1)}{3}}E_n\\
            e^{i\frac{4\pi (n-1)}{3}}E_n\\
            e^{i\frac{2\pi (n-1)}{3}}\\
            e^{i\frac{4\pi (n-1)}{3}}
        \end{pmatrix}\,.
    \end{equation}
Notice that given the parametrization of the Hamiltonian in Eq. (\ref{eq::RouterHamiltonian}) the total phase takes values in the interval $\theta \in [0, 6 \pi)$. However, the symmetries of the graph makes it sufficient to consider  $\theta \in [0, 2 \pi)$ in order to design a quantum router. Notice also that if one prepares the walker in an eigenstate, the ratio between the probability of being localized inside or outside of the loop is given by 
\begin{equation} \label{eq::probabilotyratio}
    R_n(\theta) \equiv \frac{\sum_{j=2,3,4}|\braket{j|E_n}|^2}{\sum_{j=1,5,6}|\braket{j|E_n}|^2}= E_n^2\,,
\end{equation}
which means that one may estimate the energy of the eigenstate upon estimating this probability ratio by measuring a dichotomic {\em in-out} observable.

Let us now address the main topic of the paper, i.e. how to exploit chirality to obtain directionality and, in turn, the effective routing of classical and quantum information. We start by considering a walker initially localized in one of the external sites, say $\ket{1}$, and analyze the probability of being transferred to one of the other outer 
states, i.e. $\ket{5} or \ket{6}$ at some time $t$. As we will see, it is indeed possible to achieve nearly perfect routing, that is, one may chose where to route the walker by tuning $\theta$, and obtain a transition 
probability very close to one at some time.

\begin{figure}[h!] 
	\centering
	\includegraphics[width=0.49\textwidth]{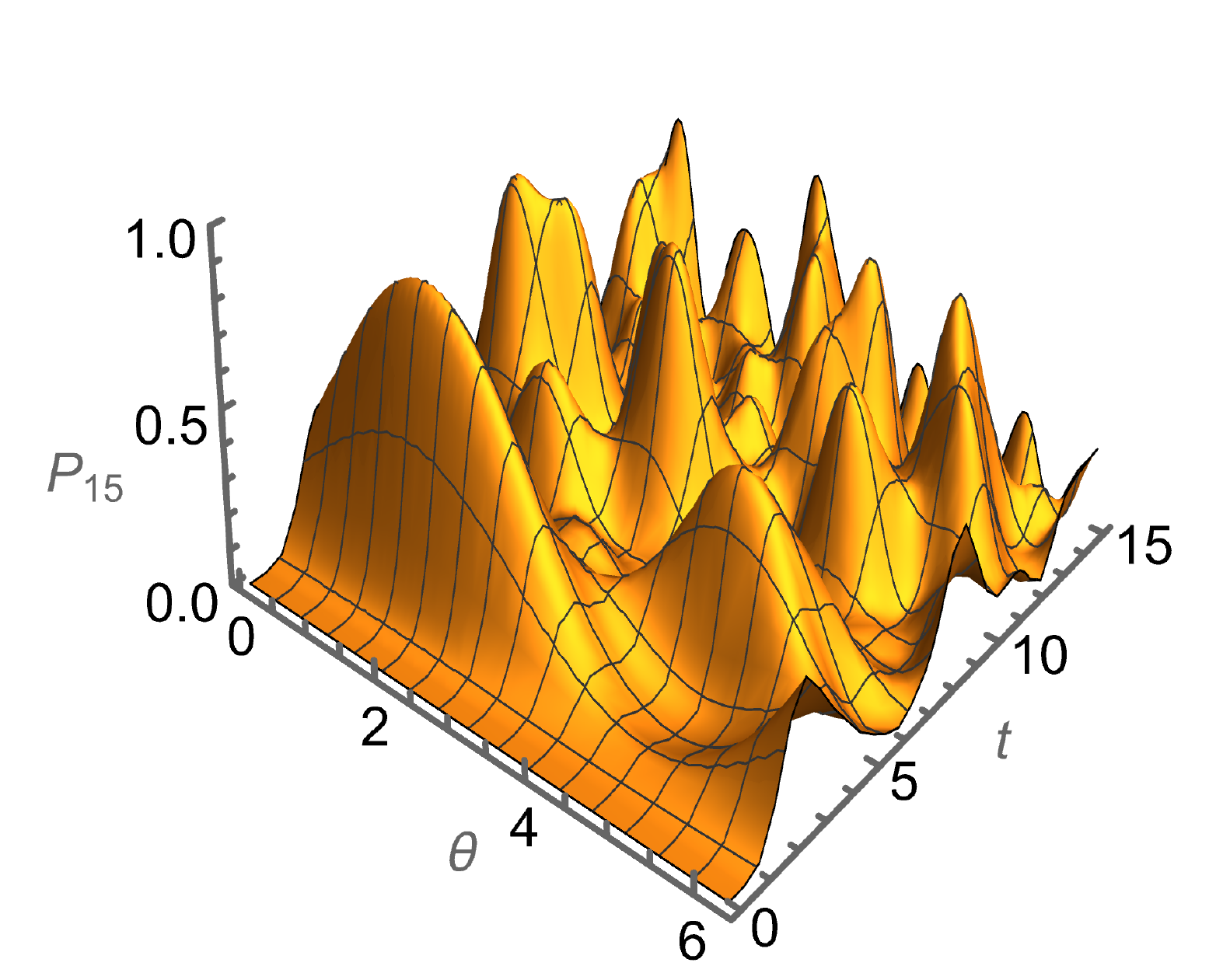}
    \includegraphics[width=0.49\textwidth]{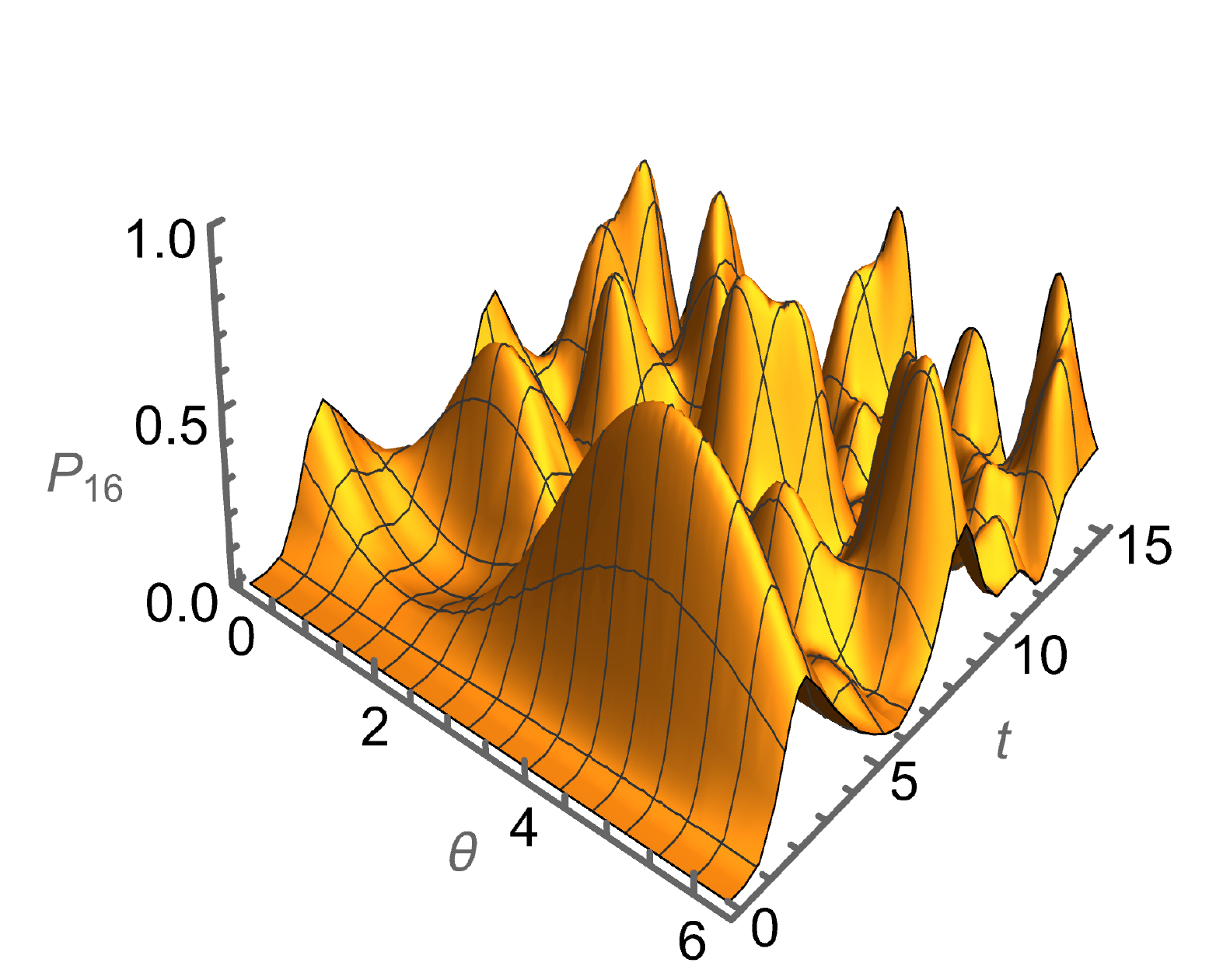}
	\caption{\label{f:f3d} The transition probabilities $P_{15}(t,\theta)$ (left) and $P_{16}(t,\theta)$ (right) as function of the interaction time $t$ and the total phase $\theta$.}
\end{figure}

Upon expanding the unitary evolution operator $U=e^{-iHt} = \sum_{n} e^{-i E_n t} \ket{E_n}\bra{E_n}$, and using 
Eqs. (\ref{eq::eigenvalues}) and (\ref{eq:eigenv}) we obtain 

    \begin{align} \label{eq::P15}
        P_{15}(t,\theta,\gamma)=\left| 
        \sum_{n=1}^6 e^{-i E_n t}\, \frac{e^{-i \frac{2\pi}{3} (n-1)}}{3 (1+E_n^2)}\right|^2 \\
        P_{16}(t,\theta,\gamma)=\left| 
        \sum_{n=1}^6 e^{-i E_n t}\, \frac{e^{-i \frac{4\pi}{3} (n-1)}}{3 (1+E_n^2)}\right|^2
    \end{align} 
Both expressions are analytic, but it is not easy to understand their general behaviour due to the high number of oscillating terms. In order to gain more insight about their properties we illustrate the 
behaviour of $  P_{15}(t,\theta) \equiv   P_{15}(t,\theta,\gamma\equiv 0) $ and $  P_{16}(t,\theta) \equiv   P_{16}(t,\theta,\gamma\equiv0) $ as a function of time and the total phase in Figure \ref{f:f3d}. As it is apparent from the plots, there are instants of time when one may switch from the situation $P_{15}\simeq 1$, $P_{16}\simeq 0$ to 
$P_{15}\simeq 0$, $P_{16}\simeq 1$ by tuning the total phase $\theta$. In other words, one may {\em route} the excitation from site 1 to site 5 or from site 1 to site 6 just by tuning the external parameter 
$\theta$ (which itself may depend on an external field). 

\begin{figure}[h!] 
	\centering
	\includegraphics[width=0.48\textwidth]{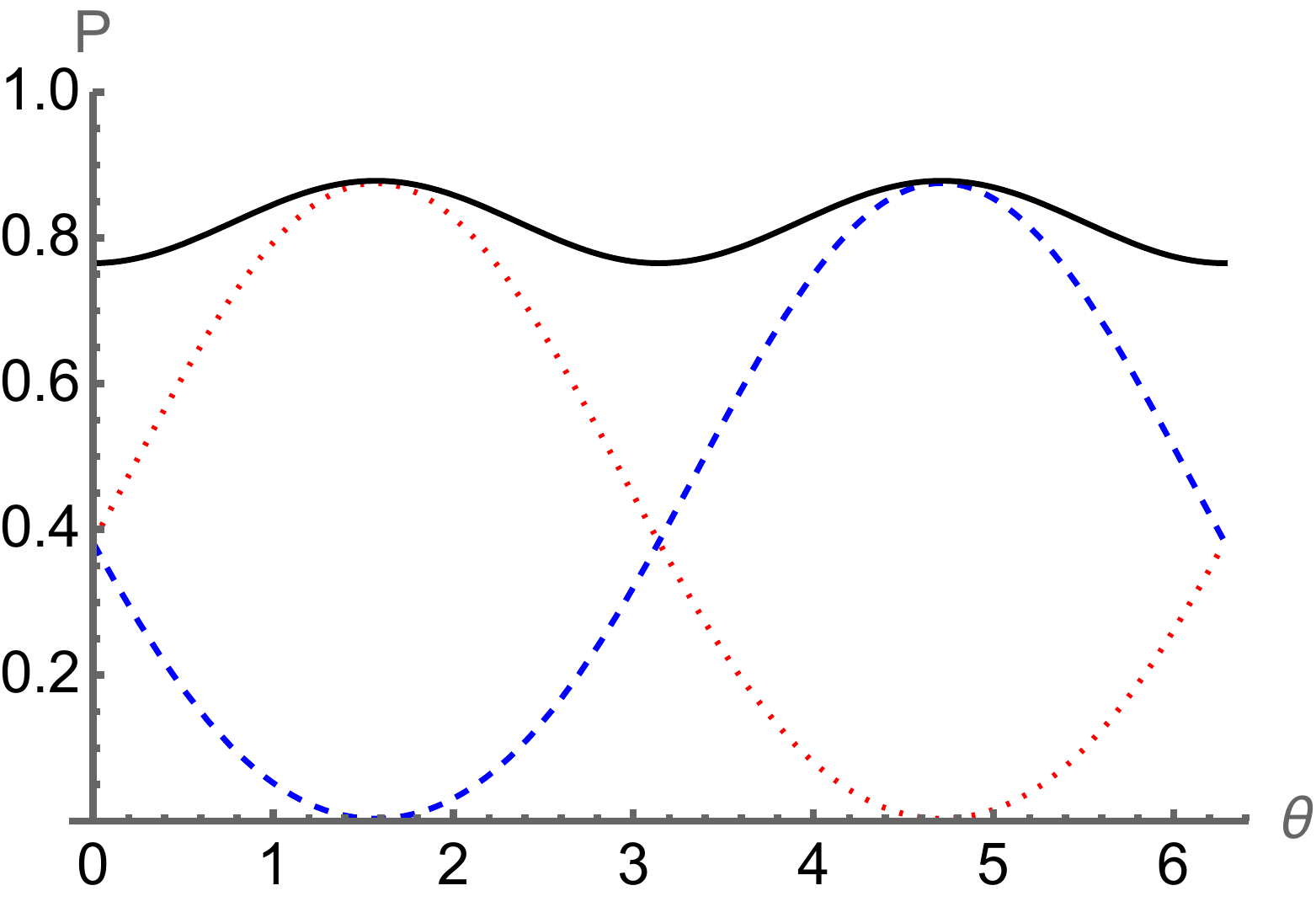}
    \includegraphics[width=0.50\textwidth]{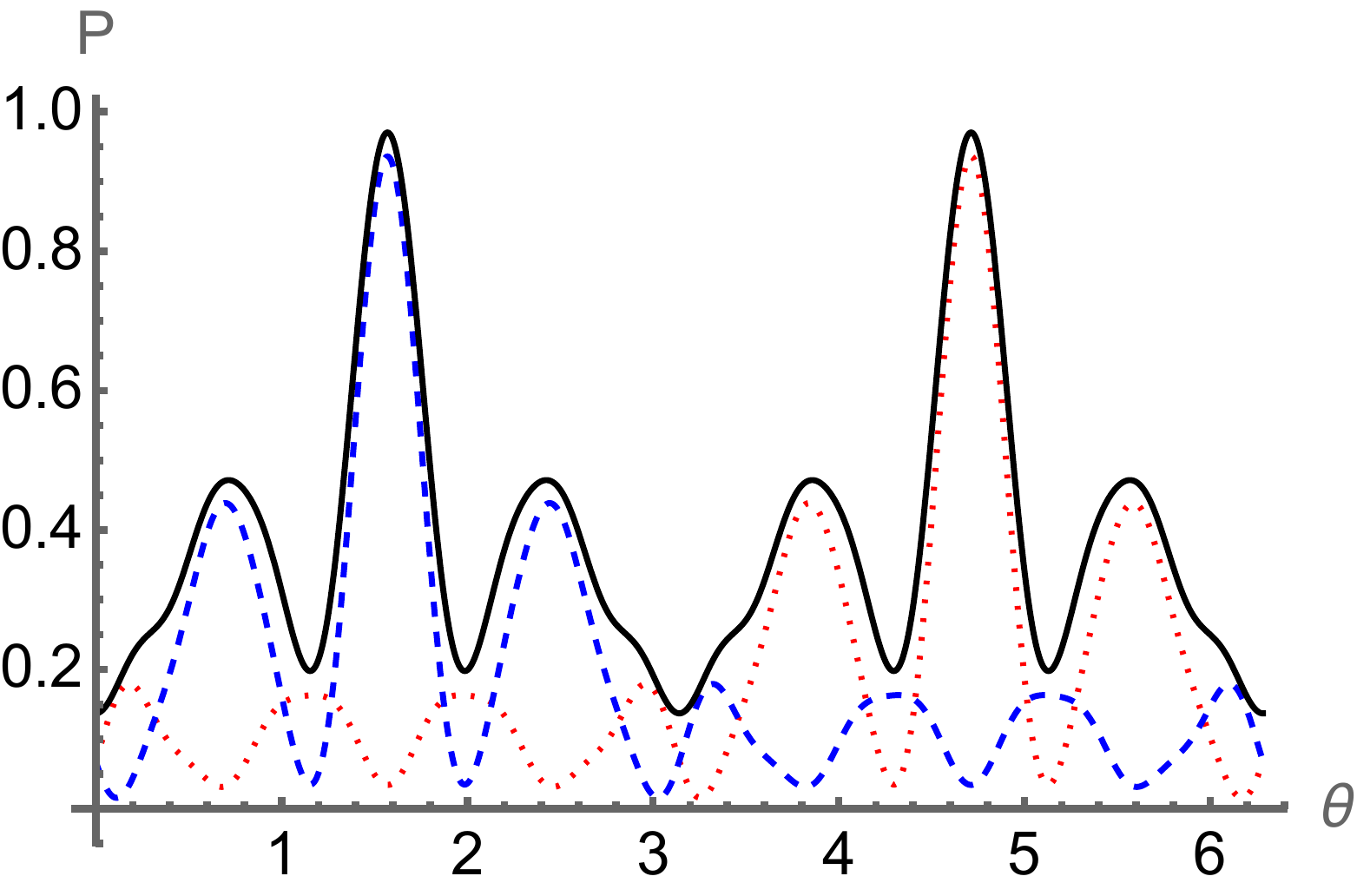}
	\caption{\label{f:f2d} The transition probabilities $P_{15}(t,\theta)$ (dotted red) 
	and $P_{16}(t,\theta)$ (dashed blue), together with their sum (solid black) as a function of 
	$\theta$ for $t=2.629$ (left) and $t=18.95$ (right). Notice that for $t=2.629$ we have nearly 
	perfect routing $1\rightarrow 5$ for $\theta=\pi/2$ and $1\rightarrow 6$ for 
	$\theta=3\pi/2$, whereas for $t=18.95$ the two phases are switched.}
\end{figure}

Upon maximizing $P_{15}(t,\theta)$ over $\theta$ at fixed $t$ (or equivalently $P_{16}$) and exploiting the symmetry $P_{15}(t,\theta)=P_{16}(t,2\pi-\theta)$ one obtains the optimal values for the routing phase, which turns out to be $\theta=\pi/2, 3/2 \pi$. This is illustrated in Fig. \ref{f:f2d}, where we show $P_{15}(t,\theta)$, $P_{16}(t,\theta)$ and their sum as a function of $\theta$ for two specific values of the interaction time. 

The routing effect is apparent from the plots. Setting the total phase to the optimal value, we have 
$P_{15} \gg P_{16}$, or vice-versa, and also have $P_{15} + P_{16}\simeq 1$, i.e. the excitation is 
effectively routed to the other outer sites, without being trapped in the loop of the graph (we have $P_{15} + P_{16}\simeq 0.88$ at $t^*=2.629$, and $P_{15} + P_{16}\simeq 0.97$ at $t^*=18.95$). It is also remarkable that the effect appears at multiple instants of time (which we refer to as $t^*$), meaning that routing may be implemented with structures and platforms of different sizes. However, for those larger values of $t^*$ the probability peaks are sharper, i.e. the routing effects is less robust. Overall, we have a tradeoff between the routing effectiveness 
and its robustness, making the first values of $t^*$ more convenient for practical applications.
Notice that for $\theta=\pi/2$ the probability $P_{15}$ may be written as $P_{15}(t,\pi/2)=\frac{4}{9}f(t)^2$ where $f(t)$ is a real function reported in Appendix A. An analogue result may be found for $P_{16}(t,\pi/2)$. 

Finally, we remark that the quantum router described in this Section has been designed setting $\gamma=0$, i.e. using only the chiral nature of the Hamiltonian and without exploiting the degree of freedom offered by the on site energies.
However, since the 
Hamiltonian in Eq. \eqref{eq::RouterHamiltonian} also depends on the diagonal elements, one may wonder about their 
potential role in routing information. Investigating the probabilities numerically, we found that by setting 
$\gamma\neq 0$ one may enhance routing (i.e. the two probabilities are closer to 1 and 0, respectively) though the optimal phase and interaction time are no longer the same. 
\begin{figure}[h!] 
	\centering
	\includegraphics[width=0.505\textwidth]{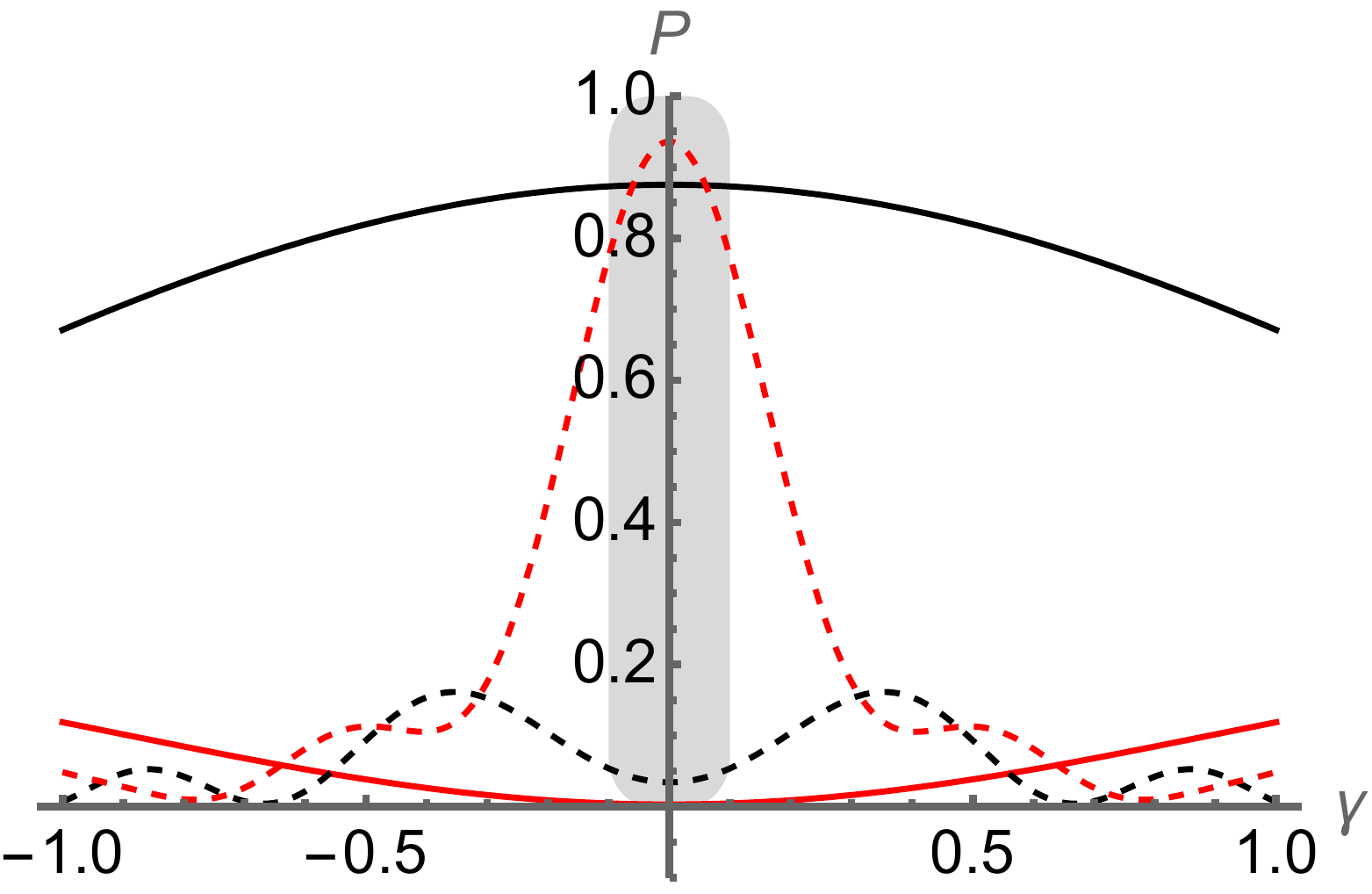}
    \includegraphics[width=0.46\textwidth]{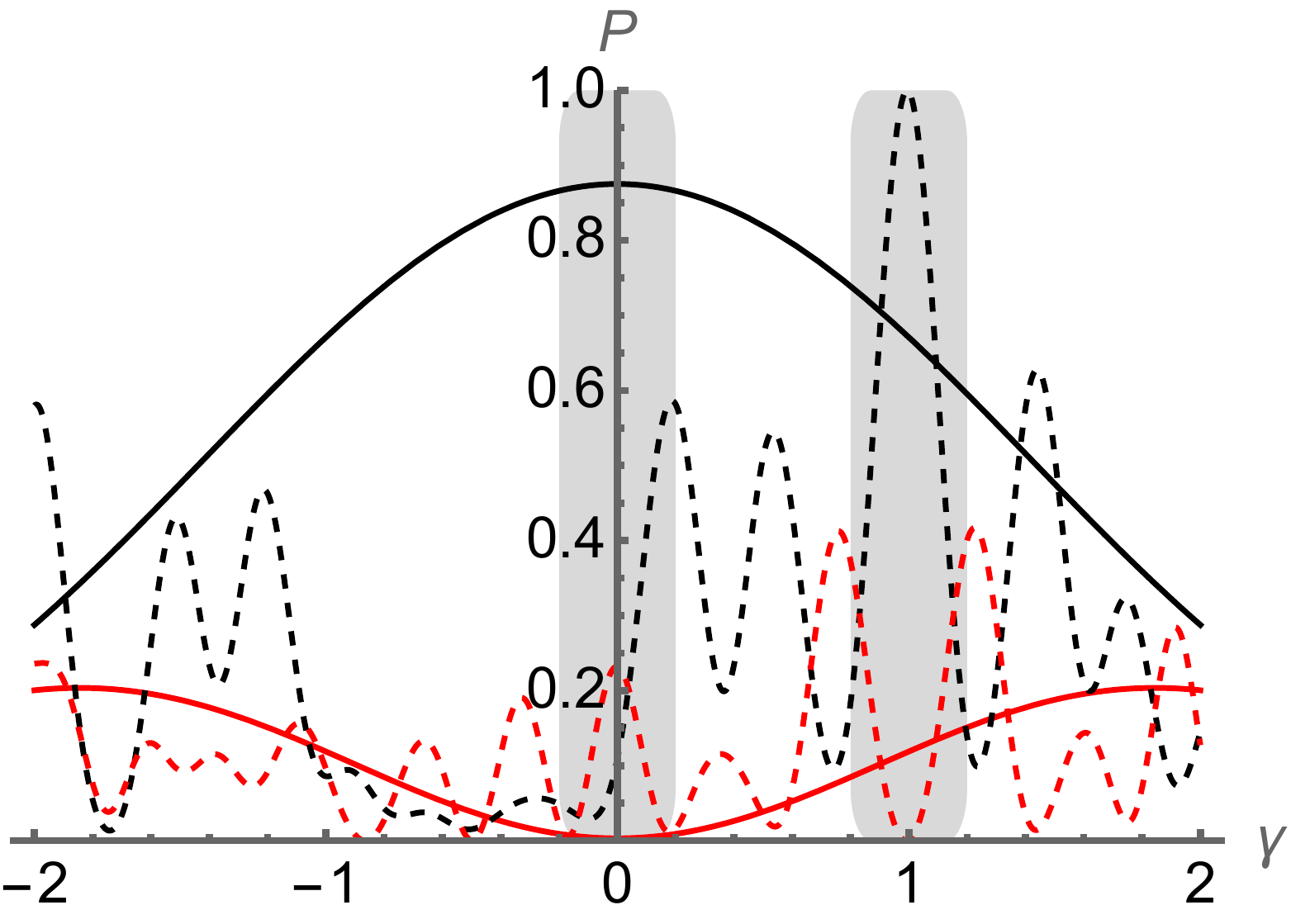}
	\caption{\label{f5:gamma}(Left): the transitions probabilities $P_{15}(t^*,\pi/2,\gamma)$ (black) and $P_{16}(t^*,\pi/2,\gamma)$ (red) as a function of $\gamma$ for $t^*=2.629$ (solid) and $t^*=18.95$ (dashed). (Right): $P_{15}(2.629,\pi/2,\gamma)$ (solid black) and $P_{16}(2.629,\pi/2,\gamma)$ (solid red) as a function of $\gamma$, 
	compared to $P_{15}(t_M,\theta_M,\gamma)$ (dashed black) and $P_{16}(t_M,\theta_M,\gamma)$ (dashed red) with $t_M\simeq 31.4$ and $\theta_M \simeq 0.11$. The gray regions highlight where the routing effect is taking place.}
\end{figure}

In other words, we have 
$P_{15}(t^*,\pi/2,\gamma)/P_{15}(t^*,\pi/2,0) < 1$ and
$P_{16}(t^*,\pi/2,\gamma)/P_{16}(t^*,\pi/2,0) > 1$ 
$\forall \gamma$, where $t^*$ is one of the optimal values of time discussed above, but we may find other values of the parameters, say $t_M$, $\theta_M$, and $\gamma_M$ for which 
$P_{15}(t_M,\theta_M,\gamma_M)/P_{15}(t^*,\pi/2,0) > 1$ and
$P_{16}(t_M,\theta_M,\gamma_M)/P_{16}(t^*,\pi/2,0) < 1$. 
The two effects are illustrated in the two panels of Fig. \ref{f5:gamma}. In the left panel, we show $P_{15}(t^*,\pi/2,\gamma)$ and $P_{16}(t^*,\pi/2,\gamma)$ as a function of $\gamma$ for the two values of $t^*$ mentioned above. As it is apparent from the plot, for $\gamma\neq 0$ the performance of the system as a router gets worse (see the region highlighted in gray). In the right panel, we instead compare $P_{15}(t^*,\pi/2,\gamma)$ and $P_{16}(t^*,\pi/2,\gamma)$ to $P_{15}(t_M,\theta_M,\gamma)$ and $P_{16}(t_M,\theta_M,\gamma)$ with $t_M\simeq 31.4$ and $\theta_M \simeq 0.11$. In this case the optimal values of $\gamma$ are different (they are $\gamma=0$ and $\gamma=1$, respectively, see the gray regions) and the $\gamma\neq 0$ case provides a more effecting routing effect.

Let us now address the routing of quantum information, i.e. of states exhibiting coherence in the site basis. 
In particular, we consider an excitation initially de-localized over the sites $\ket{1}$ and $\ket{2}$, i.e. prepared in a generic superposition of the form $$\ket{\phi_{12}}=\frac{\ket{1}+e^{i\phi}\ket{2}}{\sqrt{2}}\,,$$ and investigate the possibility of routing the state to the analogue superpositions of sites $\ket{5}$ and $\ket{3}$ or 
$\ket{6}$ and $\ket{4}$, i.e. the states 
\begin{align*}
\ket{\phi_{53}}=\frac{\ket{5}+e^{i\phi}\ket{3}}{\sqrt{2}}\\
\ket{\phi_{64}}=\frac{\ket{6}+e^{i\phi}\ket{4}}{\sqrt{2}}
\end{align*}

As we will see, routing of quantum information is indeed possible, and it is even possible to design a universal quantum router, able to route any superposition of the form $\ket{\phi_{12}}$, independenently on the value of the superposition phase $\phi$. We set $\gamma=0$ and denote the transition probabilities 
$|\braket{\phi_{53}|U|\phi_{12}}|^2$
and $|\braket{\phi_{64}|U|\phi_{12}}|^2$ by $P_R(t,\theta,\phi)$ and $P_L(t,\theta,\phi)$, respectively.
 
Using the relations
\begin{align*}
\partial_t \braket{1|U|5}&=-i\braket{1|UH|5}=-i\braket{1|U|3}\,, 
\qquad \partial_t \braket{1|U|5}=-\braket{1|HU|5}=-i\braket{2|U|5} \\
\partial_{tt}\braket{1|U|5}&=-\braket{1|HUH|5}=-\braket{2|U|3}\,,
\end{align*}
we may write
    \begin{align}
        P_R(t,\theta,\phi) = \frac14
        |\braket{1|U|5}+2i\cos\phi\partial_t\braket{1|U|5}-\partial_{tt}\braket{1|U|5}|^2\,.
    \end{align}
In particular, for $\theta=\frac{\pi}{2}$ we have  
    \begin{equation}\label{prpi2}
         P_{R}(t,\frac{\pi}{2},\phi)=\frac{1}{4}|f(t)+2i\cos\phi\partial_t f(t) -\partial_{tt} f(t)|^2\,,
    \end{equation}
where $f(t)$ is reported in Appendix A.

The dependence on the superposition phase $\phi$ is multiplied by the time derivative of the amplitude $\braket{1|U|5}$. When this derivative is zero (notice that $\braket{1|U|5}$ is a complex number) the transport probability 
is independent on $\phi$, i.e. it is the same for all the superposition states of the form $\ket{\phi_{12}}$
For those instants of time $t^*$, the value of the transport probability is 
    \begin{equation}
        P_R(t^*,\frac{\pi}{2},\phi)=\frac{1}{4}|f(t^*)-\partial_{tt} f(t)|_{t=t^*}|^2\,.
    \end{equation}
    The first value is $t^*=2.629$ and coincides with the first instant of time optiziming routing for 
    localized states. It corresponds to $ P_R(t^*,\frac{\pi}{2},\phi) = 0.83$. Upon sending $\theta$ into $2\pi -\theta$ the same line of reasoning may be applied to $P_L(t^*,\frac{\pi}{2},\phi)$. 

\begin{figure}[h!] 
	\centering
	\includegraphics[width=0.5\textwidth]{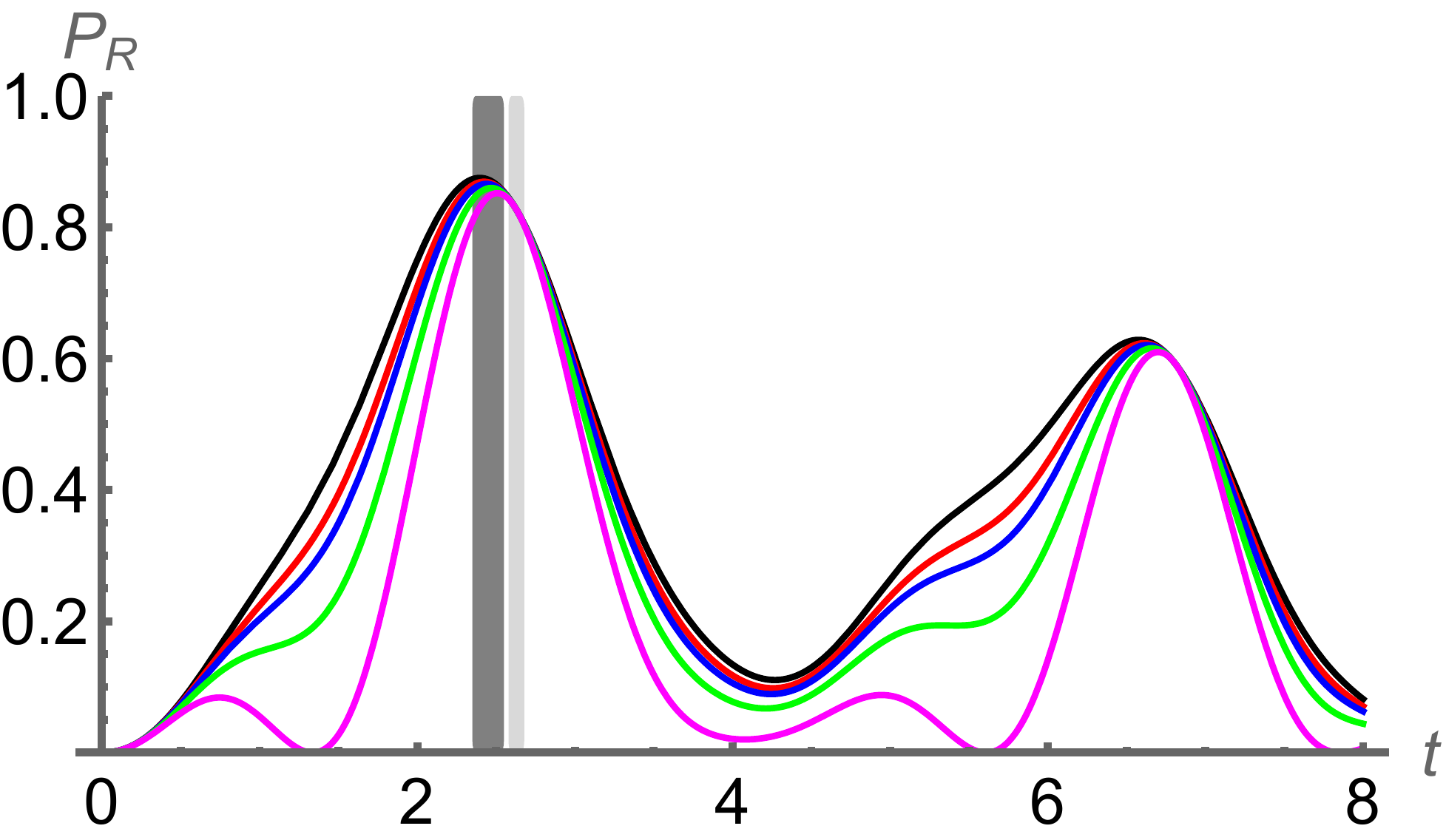}
	\caption{\label{f6:PR} The $12\rightarrow 53$ transition probability $P_R(t,\frac{\pi}{2},\phi)$ as a function of the interaction time for different values of $\phi=0,\pi/8,\pi/6,\pi/4,\pi/2$ (black, red, blue,  green, and magenta, respectively). The light gray area highlights the region around $t^*=2.629$ and the dark gray one the interval $t\in [2.35,2.55]$ (see text).}
\end{figure}

We should remark that the above setting makes the probability $\phi$-independent, but not necessarily 
optimal for all the value of $\phi$. If the value of $\phi$ is fixed, a specific optimization may be done 
separately. Notice, however, that the differences are not dramatic, and that the instants of time optimizing the routing for localized states are also {\em pretty good} for all superpositions. This is illustrated in Fig. \ref{f6:PR}, where we show $P_R(t,\frac{\pi}{2},\phi)$ for different values of $\phi$. The gray areas highlight the region around $t^*=2.629$ (light gray) where universality occurs and the interval $t\in [2.35,2.55]$ (dark gray), where the maxima for the considered values of $\phi$ are found.

Finally, we emphasize once again that chirality is a crucial ingredient to achieve the universal routing of superpositions, which would not be possible for a non-chiral CTQW, since in that case $\theta=0$.

\section{Probing the router phase}

As we have illustrated in the previous Section, the structure in Fig. \ref{fig::router}, when 
equipped with the Hamiltonian in Eq. (\ref{eq::RouterHamiltonian}), allows one to route 
classical and quantum information by tuning (and switching) $\theta$, i.e. the total 
phase that the walker acquires along the loop. The routing effect appears {\em robust} since 
the peaks in Fig. \ref{f:f2d} are {\em broad} compared to the range of variation of $\theta$. 
However, this is only a qualitative assessment, and a proper benchmark for robustness 
may be obtained only by comparing those widths to the uncertainty in determining the 
value of $\theta$ for a given physical platform. A question thus naturally arises about 
the ultimate precision achievable in estimating the total phase by probing the router. 
This problem may be addressed by the tools of quantum estimation theory 
\cite{BC94,LQE09} and quantum probing \cite{qp1,qp2,qp3,qp4,qp5,CHAPEAUBLONDEAU2022128300}. 
A natural characterization scheme is the following: the walker is repeatedly prepared 
in a known state $\ket{\psi_0}$, and is let to evolve on the graph for a given interval 
of time. Then an observable $X$ is measured, and a set of $M$ outcomes 
$\{x_1, x_2, \cdots, x_M \}$ is collected. Data are distributed according to the 
conditional distribution $p(x|\theta,t) = \left|\braket{x|U|\psi_0}\right|^2$, and the value 
of $\theta$ is inferred using an {\em estimator} $\bar\theta\equiv \bar\theta 
(\{x_1, x_2, \cdots , x_M\})$ i.e. a function of data.

\begin{figure}[h!]
\centering
\includegraphics[width=0.50\textwidth]{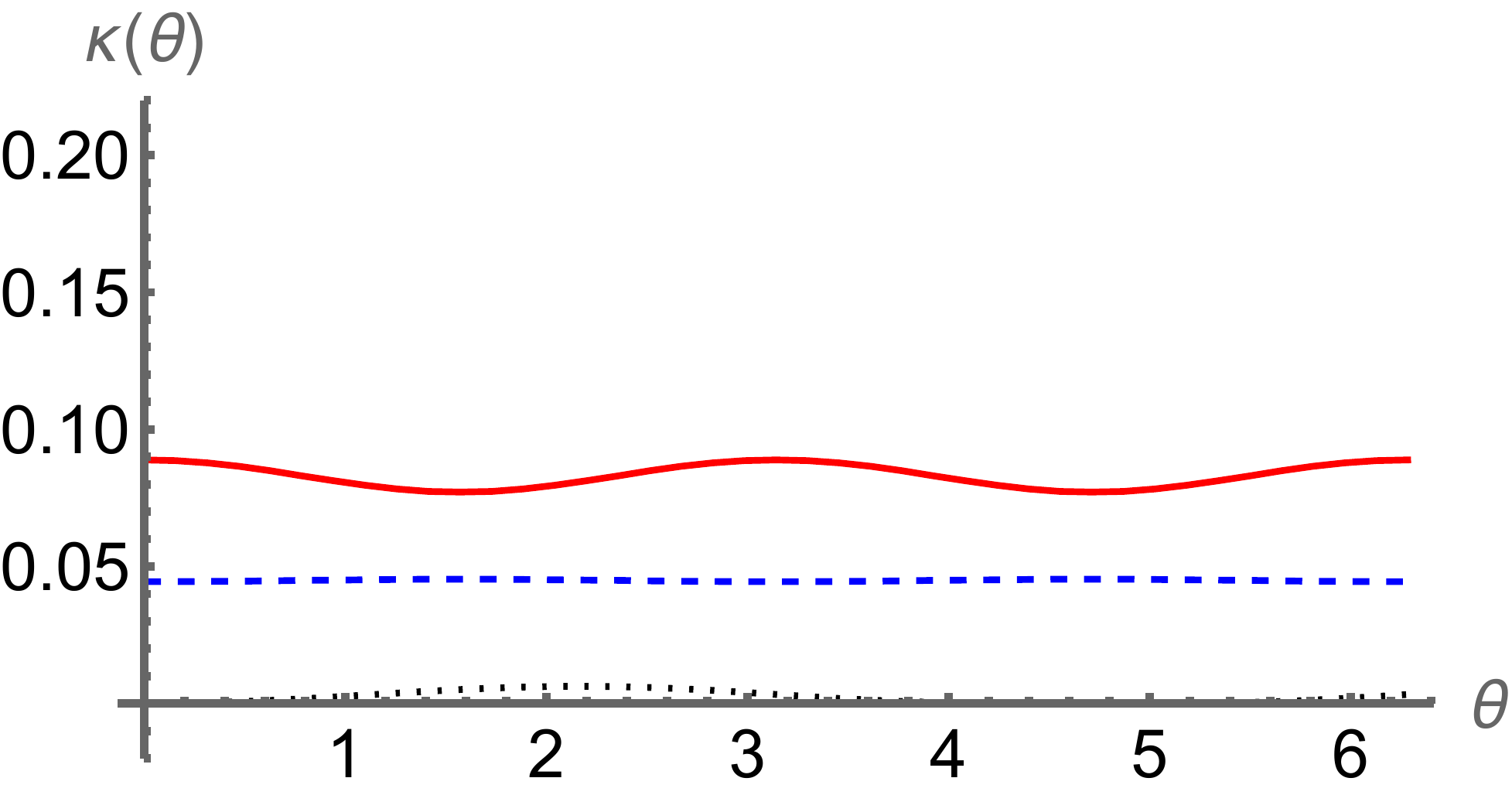}
\includegraphics[width=0.47\textwidth]{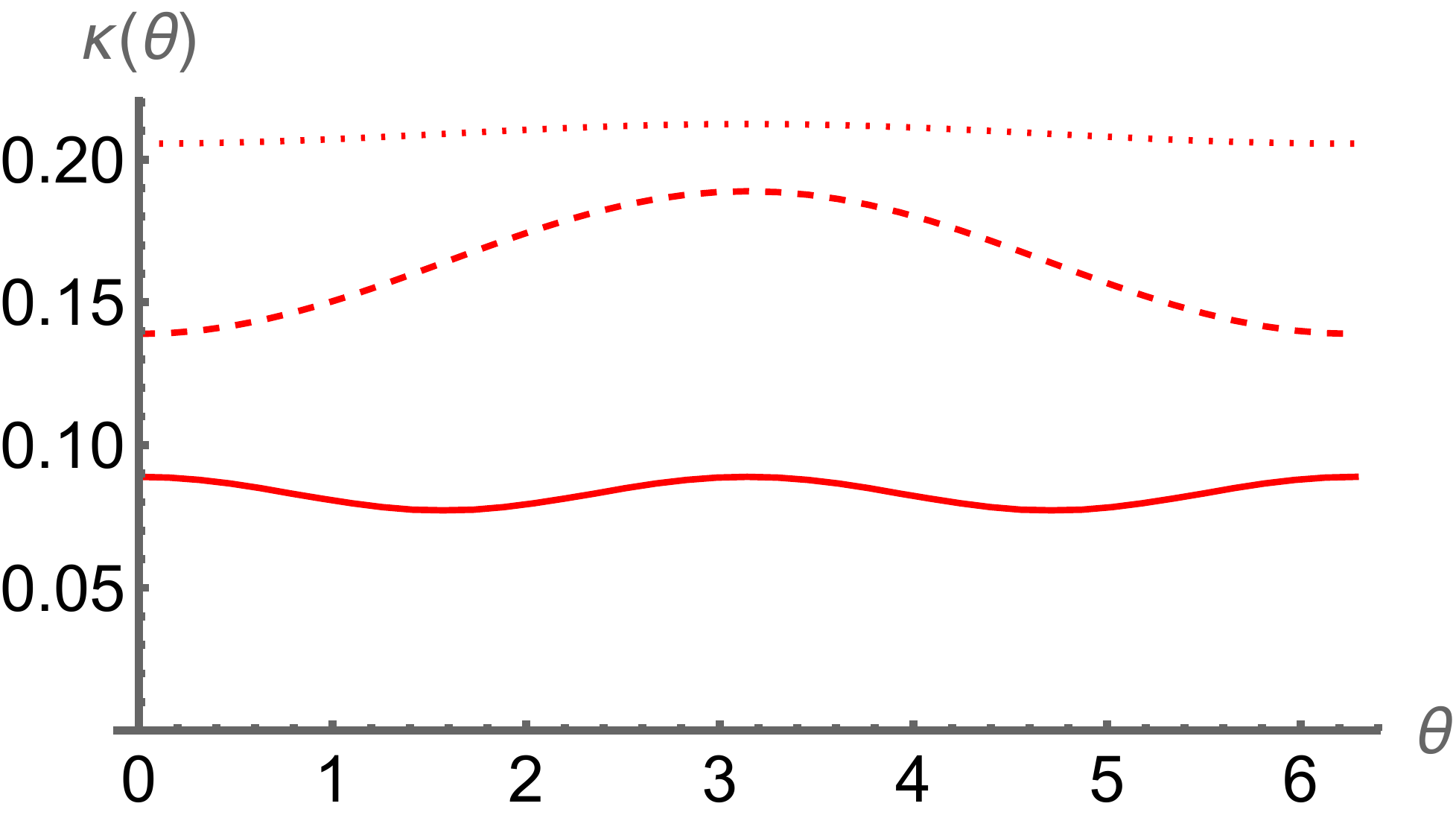}
\caption{
\label{f7:qfi} (Left): The function $\kappa (\theta) \stackrel{t\gg1}{\equiv} Q(\theta)/t^2$ as a function 
of $\theta$ for $\ket{\psi_0}=\ket{1}$ (blue dotted), $\ket{\psi_0}=\ket{2}$ (the upper red curve), and $\ket{\psi_0}=\frac{1}{\sqrt{6}}\sum_{j=1}^6\ket{j}$ (black dashed, nearly indistinguishable from the x-axes). (Right): The function $\kappa (\theta)$ as a function of $\theta$ for $\ket{\psi_0}=\ket{2}$ and $\gamma=0$ (solid), $\gamma=3$ (dashed), $\gamma=7$ (dotted).}
\end{figure}

The ultimate bound on the precision achievable by any procedure of this kind 
(i.e. the precision achievable by optimizing over all the 
observables and all the possible estimators) is bounded by the quantum 
Cram\`er-Rao bound $\hbox{Var}\bar\theta \geq [M Q(\theta)]^{-1}$, 
where $Q(\theta)$ is the quantum Fisher information (QFI) of the family of states $\ket{\psi_\theta} = U \ket{\psi_0}$ (usually referred to as a {\em quantum statistical model}). In our case the QFI is given by \cite{Liu2015} 
\begin{align}\label{qfi}
Q(\theta) = \braket{\psi_0|\Delta{\cal H}^2|\psi_0} = \braket{\psi_0|{\cal H}^2|\psi_0} - \braket{\psi_0|{\cal H}|\psi_0}^2\,,
\end{align}
where
\begin{align}
{\cal H} = -i \left( \partial_\theta U^\dag\right) U\,.
\end{align}
Notice that ${\cal H} \neq H$ since the dependence of the evolution operator $U$ on the parameter $\theta$ is non trivial (we would have ${\cal H}=H$ for and evolution operator of the form $U=e^{-i H \theta}$).
Optimized quantum probes to characterize the router are thus initial preparations $\ket{\psi_0}$ maximizing the QFI in Eq. (\ref{qfi}). Given the unitary nature of the map $\ket{\psi_0} \rightarrow \ket{\psi_\theta}$, we expect the QFI 
to grow asymptotically (i.e. for $t\gg 1$) as $Q(\theta) = \kappa(\theta)\, t^2$ independently on the initial state \cite{de2013quantum,Liu2015,can21}, with optimal probes maximizing $g(\theta)$. Actually, we have not performed an optimization over all possibile initial preparation, but rather compared the QFI obtained for some specific initial preparations, chosen because they are likely to be easily prepared on physical platforms. In the left panel of Fig. \ref{f7:qfi} we show the function $\kappa (\theta)$ as a function of $\theta$ for the localized states $\ket{\psi_0}=\ket{1}$ (blue dotted) and $\ket{\psi_0}=\ket{2}$ (the upper red curve), and for the {\em flat} superposition $\ket{\psi_0}=\frac{1}{\sqrt{6}}\sum_{j=1}^6\ket{j}$ (the lower black dashed curve, nearly indistinguishable from the x-axes). Results for the other outer and inner 
localized states are the same. The most convenient setting is thus to prepare the walker in a localized states on one of the sites of the loop and let it evolve as long as possible. 
In the right panel of Fig. \ref{f7:qfi} we illustrate the influence of the self energies, i.e. what happens for $\gamma\neq 0$. In particular, we show the behaviour of the QFI for $\ket{\psi_0}=\ket{2}$. The QFI increases and its dependence on $\theta$ weakens. For large $\gamma$ we have the asymptotic results $\kappa (\theta) = 2/9\, t^2$, which may also obtained analytically by expanding the eigenvalues and the eigenvectors of the Hamiltonian at the leading order in $\gamma$.
\section{Conclusions}
We have addressed the use of chiral quantum walks to route classical and quantum information over 
quantum network. In particular, we have shown that using a minimal graph structure is enough 
to model directional transfer of information over a network. Upon tuning the total phase along 
the loop of the graph, high-fidelity routing and transport is possible 
for localized states and coherent superpositions of them. Furthermore, we have shown how to achieve 
universal quantum routing with fidelity indipendent on the input state. In our scheme, chirality is 
governed by a single phase, and the routing probability is robust against fluctuations of this 
parameter. Finally, we have addressed characterization of quantum routers and evaluated the quantum Fisher information for different quantum probes, also showing how to exploit 
the self energies of the graph to achieve high precision in estimating the phase parameter.

Our results confirm that chiral quantum walks represent a useful resource to design quantum information protocols over networks, and pave the way to the design of larger structures to test whether chirality may be exploited to route information in multiple directions. In particular, a question arises on whether the basic triangular router analyzed in this work is still optimal for multiple routing, or larger structures with possibly more free phases are required. 
\appendix
\section{The function f(t) in Eq. (\ref{prpi2})}
The function $f(t)=f_r(t) + i f_i(t)$ is given by
\begin{align}
f_r(t) = & \,
\frac{\sin \left(\frac{\sqrt{3} t}{2}\right) \sin
   \left(\frac{\sqrt{7} t}{2}\right)}{2 \sqrt{21}}-\frac{\cos
   (t)}{6}+\frac{1}{6} \cos \left(\frac{\sqrt{3} t}{2}\right)
   \cos \left(\frac{\sqrt{7} t}{2}\right)  
   -\frac{\sin
   \left(\frac{\sqrt{3} t}{2}\right) \cos \left(\frac{\sqrt{7}
   t}{2}\right)}{2 \sqrt{3}}\nonumber \\
   & +\frac{\sin \left(\frac{\sqrt{7}
   t}{2}\right) \cos \left(\frac{\sqrt{3} t}{2}\right)}{2
   \sqrt{7}} \\
   & \nonumber \\
f_i(t) = & \,
\frac{\sin \left(\frac{\sqrt{3} t}{2}\right) \sin
   \left(\frac{\sqrt{7} t}{2}\right)}{2 \sqrt{7}}-\frac{\cos
   (t)}{2 \sqrt{3}}+\frac{\cos \left(\frac{\sqrt{3} t}{2}\right)
   \cos \left(\frac{\sqrt{7} t}{2}\right)}{2
   \sqrt{3}} -\frac{1}{2} \sin \left(\frac{\sqrt{3} t}{2}\right)
   \cos \left(\frac{\sqrt{7} t}{2}\right)\nonumber \\ 
   & +\frac{1}{2}
   \sqrt{\frac{3}{7}} \sin \left(\frac{\sqrt{7} t}{2}\right)
   \cos \left(\frac{\sqrt{3} t}{2}\right)
\end{align}
\section*{Acknowledgements}
This work has been performed under the auspices of GNFM-INdAM. MGAP thanks Claudia Benedetti, Alessandro Candeloro and Emilio Annoni for useful discussions.
\bibliography{bib_qr}

\end{document}